\documentclass[11pt,aps,preprint]{revtex4}
\usepackage[active]{srcltx}
\usepackage[utf8]{inputenc}
\usepackage{latexsym}
\usepackage{amsmath}
\usepackage{graphicx}

\begin{document}

\title{Stochastic Quantization of the Spherical Model and Supersymmetry}

\author{P. F. Bienzobaz}
\email{paulafb@if.usp.br}
\affiliation{Instituto de F\'\i sica, Universidade de S\~ao Paulo\\
Caixa Postal 66318, 05314-970, S\~ao Paulo, SP, Brazil}%

\author{Pedro R. S. Gomes}
\email{pedrorsg@fma.if.usp.br}
\affiliation{Instituto de F\'\i sica, Universidade de S\~ao Paulo\\
Caixa Postal 66318, 05314-970, S\~ao Paulo, SP, Brazil}%

\author{M. Gomes}
\email{mgomes@fma.if.usp.br}
\affiliation{Instituto de F\'\i sica, Universidade de S\~ao Paulo\\
Caixa Postal 66318, 05314-970, S\~ao Paulo, SP, Brazil}%


\begin{abstract}

We use the stochastic quantization method to construct a supersymmetric version of the quantum spherical model.
This is based on the equivalence between the Brownian motion described by
a Langevin equation and the supersymmetric quantum mechanics, which is connected with the existence of the Nicolai map.
We investigate the critical behavior of the supersymmetric model at finite as well as at zero temperatures,
showing that it exhibits both classical and quantum phase transitions,
and determine the critical dimensions.

\end{abstract}
\maketitle



\section{Introduction}

The stochastic quantization scheme was proposed by Parisi and Wu \cite{Parisi}, as an alternative method to quantize field theories.
Roughly speaking, the construction is based in an analogy between Euclidean quantum field theory whose functional integration is
weighted by the factor $e^{-\frac{S}{\hbar}}$, where $S$ is the Euclidean action,
and the Boltzmann distribution of a statistical system in equilibrium $e^{-\frac{U}{K_B T}}$, where $U$ is the energy.
By this proposal, to the quantum field it is associated a stochastic process whose evolution follows  a Langevin or Fokker-Planck equation.
The Green functions of the quantum theory are then obtained as the equilibrium limit of the correlations functions
of the stochastic theory.
Various investigations of conceptual and technical aspects of this method have been reported, as well as applications to numerical simulations in lattice field theories (see \cite{Migdal,Damgaard,Namiki1} and the references therein).

From another point of view,  stochastic quantization provides connections between field theoretical models.
Indeed, it relates a classical theory in $D$ dimensions with a theory in $D+1$ dimensions, because it introduces
an additional \textquotedblleft time\textquotedblright direction in the system, a fictitious time whose evolution is governed by the Langevin equation.
In this respect, without taking the equilibrium limit,
stochastic quantization can be interpreted as a prescription for the construction of higher dimensional field theories,
by considering the fictitious time as a physical parameter.
Interesting enough is the arising of a supersymmetry in the fictitious time direction. This is connected with the existence of the Nicolai map, which
will be discussed shortly. This issue plays a central role throughout this work.
A modern and related approach of these aspects is presented in \cite{Orlando}.

The above  observations may be positively exploited in favor of the quantization of the classical spherical model \cite{Berlin},
as explained from now on. The Hamiltonian of the model, in the presence of an external field $h$, is given by
\begin{equation}
\mathcal{H}_{c}=-\frac12\sum_{{\bf r},{\bf r}^{\prime}}
U_{{\bf r},{\bf r}^{\prime}}S_{{\bf r}}S_{{\bf r}^{\prime}}
+h\sum_{{\bf r}}S_{\bf r},
\label{SQ01a}
\end{equation}
where ${\bf r}$ and ${\bf r}'$ are lattice vectors, $\left\{S_{\bf r}\right\}$ is a
set of spin variables that assume continuous values, $-\infty<S_{\bf r}<\infty$, in a $D$-dimensional hypercubic
lattice with periodic boundary conditions; $U_{{\bf r},{\bf r}^{\prime}}$ is the interaction energy that depends only on the distance between the sites ${\bf r}$ and ${\bf r}'$, $U_{{\bf r},{\bf r}'}\equiv U(|{\bf r}-{\bf r}'|)$.
The $S_{\bf r}$ variables are subject to the spherical constraint,
\begin{equation}
\sum_{\bf r} S_{\bf r}^2=N,
\label{SQ02b}
\end{equation}
where $N$ is the total number of lattice sites.
The first step towards the quantization of the model  is to
provide a dynamic to the system. This can be done by including in the above Hamiltonian a term involving the
conjugated momenta of the spin variables or,
equivalently, by adding to its Lagrangian a kinetic term containing time derivatives of the spin variables.
However, there is no unique way to choose the form of these terms and, moreover, different forms yield
different dynamical behavior \cite{Obermair,Nieuwenhuizen,Vojta1}.
The stochastic quantization furnishes a natural framework to ameliorate this situation, because
the spin variables are assumed to satisfy the Langevin equation, which then  gives rise to a dynamic for the system, with
the identification of the fictitious time as the ordinary time.
We wish to apply this formalism to construct a supersymmetric version of the quantum spherical model.
As we will see, the result of this process is naturally an off-shell supersymmetric theory, which is possible, as we said, due to the existence of the Nicolai map underlying to it.

Another question is about the critical behavior of the resulting supersymmetric model.
In other words, we intend to investigate the existence of classical and quantum phase transitions, namely, phase transitions that occurs at
finite and at zero temperatures, respectively, as the nonsupersymmetric counterpart
exhibits both of them.
An early supersymmetric formulation (on-shell) of the quantum spherical model was constructed in \cite{Gomes}, in which was also
verified the existence of both classical and quantum phase transitions whenever the supersymmetry was broken.

Our work is organized as follows. We start in section II by reviewing some aspects of stochastic quantization
in quantum field theory and in quantum mechanics, which are important for discussions in the subsequent sections.
Section III is  devoted to the construction of a supersymmetric quantum spherical model from the
classical one via stochastic quantization. In section IV we use the saddle point method to evaluate the partition function of the supersymmetric model.
The critical behavior at finite and at zero temperatures is discussed in section V. A summary and additional comments are presented in section VI. In the Appendix, we discuss the stochastic quantization of the mean spherical model.


\section{Some Remarks on Stochastic Quantization}

Before  considering the stochastic quantization of the spherical model, let us initially outline
some basic steps of the prescription for the stochastic quantization of a scalar field. Then we discuss
the equivalence between Langevin equation and supersymmetric quantum mechanics. This simple analysis is an important
point of our discussions to be performed in the next sections.

\subsection{Stochastic Quantization of a Scalar Field Theory}\label{section2a}

In a general setting, the stochastic quantization of a field theory model, with a scalar field $\phi(x)$ depending on
$D$-dimensional Euclidean space-time coordinates $x$ and specified by an action $S$, proceeds as follows. The field $\phi(x)$ is assumed to depend on a
fictitious time $\tau$, $\phi(x)\rightarrow\phi(x,\tau)$, such that the evolution in this variable is governed by the
Langevin equation
\begin{eqnarray}
\frac{\partial \phi(x,\tau)}{\partial \tau}
=-\frac{\kappa}{2}\frac{\delta S}{\delta \phi(x,\tau)}+\eta(x,\tau),
\label{langevin}
\end{eqnarray}
where $\kappa$ is the diffusion coefficient that can be conveniently chosen,
$\eta(x,\tau)$ is a noise with a Gaussian distribution,
\begin{eqnarray}
\left<\eta(x,\tau)\right>=0,~~~
\left<\eta(x,\tau)\eta(x',\tau')\right>
=\kappa\,\delta(\tau-\tau')\delta^D(x-x'),
\end{eqnarray}
and $S$ is the Euclidean action  that must include also an integration over the fictitious time,
\begin{equation}
S=\int d^Dxd\tau\mathcal{L}(\phi(x,\tau),\partial_{\mu}\phi(x,\tau)).
\end{equation}
The theory arising from this process is a $D+1$-dimensional one.
Let us describe how to come back to a $D$-dimensional but quantum theory.
For a specified initial condition in the fictitious time,
the Langevin equation (\ref{langevin})   has the noise-dependent solution $\phi_{\eta}(x,\tau)$ whose
correlation functions are  defined as
\begin{equation}
\langle \phi_{\eta}(x_1,\tau_1)\cdots\phi_{\eta}(x_N,\tau_N)\rangle_{\eta}\equiv
\int \mathcal{D}\eta~ \phi_{\eta}(x_1,\tau_1)\cdots\phi_{\eta}(x_N,\tau_N) P[\eta],
\end{equation}
with the normalized probability
\begin{equation}
P[\eta]=\frac{1}{Z}\exp\left[-\frac{1}{2\kappa}
\int d^Dx d\tau~\eta^2(x,\tau)\right]
\end{equation}
and the partition function $Z$ given by
\begin{eqnarray}
Z=\int\mathcal{D}\eta \exp\left[-\frac{1}{2\kappa}
\int d^Dx d\tau~\eta^2(x,\tau)\right].
\end{eqnarray}
The equivalence of this procedure with other  quantization methods  is established at the level of the Green functions \cite{Parisi}.
In fact, it has been shown that under certain conditions which include the semi-positivity of the associated Fokker-Planck Hamiltonian, if we take
$\tau_1=\cdots=\tau_N\equiv\tau$ and then take the equilibrium limit $\tau\rightarrow\infty$, we obtain the
time-ordered Green functions of the quantized field theory,
\begin{equation}
\lim_{\tau\rightarrow\infty}\langle \phi(x_1,\tau)\cdots\phi(x_n,\tau)\rangle_{\eta}=\langle 0|T\phi(x_1)\cdots\phi(x_N) |0\rangle.
\end{equation}
The quantization procedure above described can be summarized as follows.
We start with an action in $D$-dimensional Euclidean space-time. The field is
assumed to depend on a new time coordinate so that the resulting theory is a $D+1$-dimensional one. In order to
reach the quantum theory we take the equilibrium limit so eliminating the fictitious time and returning
to a $D$-dimensional situation.

However, as we argued before, we may take a different route and consider this process as a prescription for the construction of $D+1$-dimensional theories
from $D$-dimensional ones. In this work we are interested in the $D+1$-dimensional emerging theory itself.
In quantum mechanics, for example, this establishes a quite interesting connection between the Brownian motion and
the supersymmetric quantum mechanics.

\subsection{Langevin Equation and Supersymmetric Quantum Mechanics}

Let us start by considering a point particle whose spatial coordinate is $x$.
Following the stochastic quantization prescription, the  coordinate $x$ is assumed to depend on the
fictitious time $\tau$, $x\rightarrow x(\tau)$. As mentioned before, we will not
eliminate the fictitious time in the end of the process, but
instead we will identify it just as  the ordinary time.
The Langevin equation (with $\kappa=1$),
\begin{equation}
\frac{dx(\tau)}{d\tau}=-\frac12\frac{\delta S}{\delta x}+\eta(\tau),
\label{l1a}
\end{equation}
describes the Brownian motion of a particle in a heat bath subjected to a random noise.
The partition function is given by the integral over the noise,
\begin{equation}
Z=\int\mathcal{D}\eta\exp\left[-\frac{1}{2}\int d\tau~\eta^2(\tau)\right].
\end{equation}
We perform a change of variable of integration $\eta(\tau)\rightarrow x(\tau)$ defined by the
Langevin equation (\ref{l1a}) to write the partition function as
\begin{eqnarray}
Z= \int\mathcal{D}x
\det\left[\frac{\delta \eta(\tau)}{\delta x(\tau')}\right]
\exp\left\{-\frac{1}{2}\int d\tau\left(
\frac{d x(\tau)}{d\tau}+
\frac12\frac{\delta S}{\delta x}\right)^2\right\}.
\end{eqnarray}
The determinant can be written as an integral over anticommuting (Grassmann) variables $\psi$ and~$\bar\psi$,
\begin{eqnarray}
\det\left[\frac{\delta\eta(\tau)}{\delta x(\tau')}\right]
=\int\mathcal{D}\psi\mathcal{D}\bar{\psi}\exp\left[
\int d\tau~\bar{\psi}\left(\frac{d}{d\tau}
+V^{\prime}\right)\psi\right],
\end{eqnarray}
where we have introduced $V\equiv\frac12\frac{\delta S}{\delta x}$ with $V^{\prime}=\frac{\partial V}{\partial x}$.
We then write the partition function
\begin{eqnarray}
Z=\int\mathcal{D}x\mathcal{D}\psi\mathcal{D}\bar{\psi}
\exp\left\{-\int d\tau\left[\frac12\left(\frac{dx}{d\tau}\right)^2+
\frac12 V^2-\bar{\psi}\left(\frac{d}{d\tau}+V^{\prime}\right)\psi\right]\right\},
\end{eqnarray}
where we have discarded surface terms. From this expression, we may read out the effective Euclidean Lagrangian
\begin{eqnarray}
\mathcal{L}=\frac12\left(\frac{dx}{d\tau}\right)^2+
\frac12 V^2-\bar{\psi}\left(\frac{d}{d\tau}+V^{\prime}\right)\psi,
\end{eqnarray}
that is exactly the Lagrangian of the supersymmetric quantum mechanics \cite{Nicolai1,Witten}. As can be easily verified, it
is invariant under the supersymmetry transformations
\begin{equation}
\delta_{\epsilon}x=-\bar{\psi}\epsilon,
~~~\delta_{\epsilon}\psi=\dot{x}\,\epsilon
-V\,\epsilon~~~\text{and}~~~\delta_{\epsilon}\bar{\psi}=0
\end{equation}
and
\begin{equation}
\delta_{\bar\epsilon}x=\bar{\epsilon}\psi,~~~\delta_{\bar\epsilon}\psi=0~~~\text{and}~~~\delta_{\bar\epsilon}\bar{\psi}=\bar{\epsilon}\,\dot{x}
+\bar{\epsilon}\,V.
\end{equation}
As usual, this supersymmetry can be cast in an off-shell formulation by introducing an auxiliary field $F$,
so that the Lagrangian becomes
\begin{eqnarray}
\mathcal{L}=\frac12\left(\frac{dx}{d\tau}\right)^2-\frac12 F^2+
 F V-\bar{\psi}\left(\frac{d}{d\tau}+V^{\prime}\right)\psi.
\end{eqnarray}
From the off-shell supersymmetry it is straightforward to construct the superspace formalism, which for the quantum mechanics
it is constituted by the anticommuting coordinates $\theta$ and $\bar\theta$ besides the time $\tau$.
For example, the superfield is expanded in terms of $\theta$ and $\bar\theta$,
\begin{equation}
\Phi=x+\bar\psi\theta+\bar\theta\psi+\bar\theta\theta F.
\end{equation}
With the supercovariant derivatives
\begin{equation}
D\equiv \frac{\partial}{\partial \bar\theta}+\theta\frac{\partial}{\partial \tau}~~~\text{and}~~~
\bar D\equiv  \frac{\partial}{\partial\theta}+\bar\theta\frac{\partial}{\partial \tau},
\end{equation}
we may write the superspace action from the Lagrangian
\begin{equation}
\mathcal{L}=\frac12\bar D\Phi D\Phi+L(\Phi)~~~\Rightarrow~~~S=\int d\tau d\theta d\bar\theta\, \mathcal{L},
\end{equation}
such that $L^{\prime}\equiv V$. In general, the existence of local or global symmetries implies relations
between the Green functions of the corresponding theory, which are generically known as Ward identities.
In particular, that is the case for the supersymmetry. A natural question then is about the role of the supersymmetry
in the case discussed here.
Interesting enough is the fact that this supersymmetry is the reflex of
the forward and backward time propagation symmetry in an equivalent Fokker-Planck formulation, corresponding to the occurrence of two
Fokker-Planck Hamiltonians known in the literature as forward and backward Hamiltonians  \cite{Gozzi}.
Nevertheless, in the equilibrium limit $\tau\rightarrow\infty$ only the Hamiltonian associated with forward
time propagation will be relevant.

The procedure outlined above is related with the existence of the so called Nicolai map \cite{Nicolai}. The Nicolai map
deals with the characterization of supersymmetric theories by means of functional integration measures.
In general terms, if we start with a supersymmetric theory and then integrate out the fermionic fields from the path integral, this will
produce a non-trivial determinant. There exists a transformation (the Nicolai map) for the bosonic fields whose Jacobian
cancels exactly the determinant coming from the fermionic integration. The result is just a bosonic free action.
In the quantum mechanical case, the Langevin equation constitutes an explicit realization of the Nicolai map!
The route from Langevin equation towards the supersymmetric quantum mechanics corresponds to a Nicolai mapping in the inverse direction.



\section{Supersymmetric Quantum Spherical Model}

Based on the existence of the Nicolai map,
in this section we will obtain the supersymmetric quantum version of the spherical model from
the stochastic quantization procedure of its classical counterpart.
We will not consider
a dynamical model from the beginning. In fact, as we mentioned in the Introduction, there is no necessity of the addition of a kinetic term.
On the contrary, the dynamics arises from stochastic quantization scheme as long as we recognize the fictitious time as the ordinary time.
For the strict spherical model, to be discussed here, we need to implement the stochastic quantization for constrained systems \cite{Namiki,Justin},
in which the Langevin equation itself is considered as a constraint.

Let us start with the classical Hamiltonian of the spherical model without external field
\begin{equation}
\mathcal{H}_{c}=-\sum_{{\bf r},{\bf r}^{\prime}}
U_{{\bf r},{\bf r}^{\prime}}S_{{\bf r}}S_{{\bf r}^{\prime}},
\label{SQ01}
\end{equation}
under the spherical constraint
\begin{equation}
\sum_{\bf r} S_{\bf r}^2=N.
\label{SQ02}
\end{equation}
For simplicity, we absorbed the factor $\frac12$ in the interaction energy $U_{{\bf r},{\bf r}^{\prime}}$.
The spin variables are assumed to depend on the fictitious time that we designate by $t$, $S_{\bf r}\rightarrow S_{\bf r}(t)$, and
the action is given by
\begin{eqnarray}
S=\int dt\, \mathcal{L}=\int dt \left(-\sum_{{\bf r},{\bf r}'}U_{{\bf r},{\bf r}'}
S_{{\bf r}}S_{{\bf r}'}+\sigma\left(\sum_{{\bf r}}S_{\bf r}^2-N\right)\right).
\label{SQ03}
\end{eqnarray}
In the functional approach, by integrating over the auxiliary field $\sigma$ means that we are dealing with the strict spherical model and not the
mean spherical model. The Langevin equation is written as
\begin{eqnarray}
\frac{dS_{{\bf r}}}{dt}=
-\frac12\frac{\delta S}{\delta S_{{\bf r}}}+\eta_{{\bf r}}(t),
\label{SQ04}
\end{eqnarray}
where $\eta_{\bf r}$ is the Gaussian noise of each lattice site and we are setting $\kappa=1$.
After calculating the functional derivative of the action with respect to the $S_{{\bf r}}$ variable,
the Langevin equation takes the form
\begin{eqnarray}
\dot{S}_{{\bf r}}
=\sum_{{\bf r}'}U_{{\bf r},{\bf r}'}S_{{\bf r}'}
-\sigma S_{{\bf r}}+\eta_{{\bf r}}.
\label{SQ05}
\end{eqnarray}
It is convenient to introduce a specific notation
to unify the Langevin and the constraint equations \cite{Brunelli}.
To this end, first we consider the set of $N+1$ fields $(S_{\bf r},\sigma)$ as components of a variable
$\mathcal{S}_a$:
\begin{eqnarray}
\mathcal{S}_a\equiv\left\{
\begin{array}
[c]{cc}%
S_{{\bf r}}, ~~~a=1,2,\cdots,N\\
\!\!\!\!\!\!\!\!\!\!\sigma, ~~~~a=N+1
\end{array}
\right.,
~~~~~
\mathcal{N}_a\equiv\left\{
\begin{array}
[c]{cc}%
\eta_{{\bf r}}, ~~~a=1,2,\cdots,N\\
\!\!\!\!\!\!\!\!\!0, ~~~~a=N+1
\end{array}
\right..
\label{SQ010}
\end{eqnarray}
Next, we introduce the functions
\begin{eqnarray}
\mathcal{F}_a(\mathcal{S}_a)\equiv\left\{
\begin{array}
[c]{cc}%
F_{{\bf r}}(S_{\bf r},\sigma), ~~~a=1,2.\cdots, N\\
\!\!\!\!\!\!\!\!F(S_{\bf r}), ~~~~a= N+1
\end{array}
\right.,
\label{SQ09}
\end{eqnarray}
with the definitions
\begin{eqnarray}
F_{{\bf r}}(S_{\bf r},\sigma)\equiv\dot{S}_{{\bf r}}
-\sum_{{\bf r}'}U_{{\bf r},{\bf r}'}S_{{\bf r}'}
+\sigma S_{{\bf r}}
\label{SQ06}
\end{eqnarray}
and
\begin{eqnarray}
F(S_{\bf r})\equiv\sum_{{\bf r}}S_{{\bf r}}^2-N.
\label{SQ07}
\end{eqnarray}
According to this notation, the Langevin and the constraint equations are unified into the equation
$\mathcal{F}(\mathcal{S})=\mathcal{N}$.
The partition function is given by the integral over the noise
\begin{eqnarray}
Z=\int \mathcal{D}\mathcal{N}~
\text{e}^{-\frac12\int dt\sum_{a}\mathcal{N}_{a}^2}.
\label{SQ08}
\end{eqnarray}
The integration measure $\mathcal{D}\mathcal{N}$ symbolically stands for the functional integration over all the site lattices,
$\mathcal{D}\mathcal{N}\equiv\prod_{a}\mathcal{D}\mathcal{N}_{a}=\prod_{\bf r}\mathcal{D}\eta_{\bf r}$.
By introducing an unity in $Z$, we have
\begin{equation}
Z=\int(\mathcal{D}\mathcal{S})(\mathcal{D}\mathcal{N})
\det\left(\frac{\delta \mathcal{F}_a}{\delta \mathcal{S}_b}\right)
\delta(\mathcal{F}_a(\mathcal{S})-\mathcal{N}_a)\,\text{e}^{-\frac12\int dt\sum_a\mathcal{N}_a^2}.
\label{SQ011}
\end{equation}
The determinant can be expressed in terms of integrals over fermionic fields
$\Psi_a$ and $\bar\Psi_a$ as
\begin{eqnarray}
\det\left(\frac{\delta \mathcal{F}_a}{\delta \mathcal{S}_b}\right)
=\int(\mathcal{D}{\Psi})(\mathcal{D}\bar\Psi)
\exp\left[\int dt dt'\sum_{a,b}
\bar{\Psi}_a\frac{\delta \mathcal{F}_a}{\delta \mathcal{S}_b}\,
\Psi_b\right],
\label{SQ012}
\end{eqnarray}
with
\begin{eqnarray}
\Psi_a\equiv\left\{
\begin{array}
[c]{cc}%
\psi_{{\bf r}}, ~a=1,2,\cdots,N\\
\!\!\!\!\!\!\!\!\zeta, ~~~~~~a=N+1
\end{array}
\right.
~~~\text{and}~~~
\bar{\Psi}_a\equiv\left\{
\begin{array}
[c]{cc}%
\bar{\psi}_{{\bf r}}, ~a=1,2,\cdots,N\\
\!\!\!\!\!\!\!\!\!\bar{\zeta}, ~~a=N+1
\end{array}
\right.
\label{SQ016}
\end{eqnarray}
and the delta function as
\begin{eqnarray}
\delta(\mathcal{F}_a-\mathcal{N}_a)=\int(\mathcal{D}\Lambda)~
\text{e}^{-\int dt\sum_a\Lambda_a(\mathcal{F}_a-\mathcal{N}_a)},
\label{SQ013}
\end{eqnarray}
with
\begin{eqnarray}
\Lambda_a\equiv\left\{
\begin{array}
[c]{cc}%
\lambda_{\bf r}, ~~~a=1,2,\cdots, N\\
\!\!\!\!\!\!\!\!\alpha, ~~~~a= N+1.
\end{array}
\right.
\end{eqnarray}
So the partition function acquires the form
\begin{eqnarray}
Z&=&\int (\mathcal{D}\mathcal{N})(\mathcal{D}\mathcal{S})
(\mathcal{D}\Lambda)(\mathcal{D}\Psi)(\mathcal{D}\bar{\Psi})\nonumber\\
&\times& \exp\left\{-\int dt\sum_a\left[\frac12\mathcal{N}_a^2+
\Lambda_a(\mathcal{F}_a-\mathcal{N}_a)\right]+\int dt dt'
\sum_{a,b}\bar{\Psi}_a\frac{\delta \mathcal{F}_a}{\delta \mathcal{S}_b}\,\Psi_b\right\}.
\label{SQ014}
\end{eqnarray}
By integrating over the variable $\mathcal{N}$  and using the fact that $\mathcal{N}_a=0$ for $a=N+1$, we have
\begin{eqnarray}
Z=\int(\mathcal{D}\mathcal{S})
(\mathcal{D}\Lambda)(\mathcal{D}\Psi)(\mathcal{D}\bar{\Psi})
\exp\left[\int dt\sum_{a=1}^N\frac12\Lambda_a^2
-\int dt \sum_a\Lambda_a\mathcal{F}_a
+\int dt dt'\sum_{a,b}\bar{\Psi}_a\frac{\delta \mathcal{F}_a}{\delta \mathcal{S}_b}\,\Psi_b\right].
\label{SQ015}
\end{eqnarray}
This result is the master functional and  when written in terms of the components it becomes
\begin{eqnarray}
Z=\int (\mathcal{D}S)(\mathcal{D}\lambda)
(\mathcal{D}\sigma)(\mathcal{D}\alpha)(\mathcal{D}{\psi})
(\mathcal{D}\bar\psi)(\mathcal{D}{\zeta})
(\mathcal{D}\bar\zeta)~\text{e}^{-S_{mast}},
\label{SQ017}
\end{eqnarray}
with the master action
\begin{eqnarray}
S_{mast}&=&\int dt\left[-\frac12\sum_{{\bf r}}\lambda_{{\bf r}}^2
+\sum_{{\bf r}}\lambda_{{\bf r}}
\dot{S}_{{\bf r}}-
\sum_{{\bf r},{\bf r}'}U_{{\bf r},{\bf r}'}\lambda_{\bf r}S_{{\bf r}'}
+\sigma \sum_{\bf r}\lambda_{\bf r}S_{{\bf r}}+\alpha\left(
\sum_{{\bf r}}S_{{\bf r}}^2-N\right)\right.
\nonumber\\
&-&\left.\sum_{{\bf r}}
\bar{\psi}_{{\bf r}}\dot{\psi}_{{\bf r}}
+\sum_{{\bf r},{\bf r}'}U_{{\bf r},{\bf r}'}\bar{\psi}_{{\bf r}}
\psi_{{\bf r}'}
-\sigma\sum_{\bf r}\bar{\psi}_{{\bf r}}\psi_{{\bf r}}- \sum_{{\bf r}}
\bar{\psi}_{{\bf r}}\,S_{{\bf r}}\,\zeta
-2\sum_{\bf r}\bar{\zeta}\,S_{{\bf r}}\,\psi_{{\bf r}}  \right].
\label{SQ018}
\end{eqnarray}
By integrating over the fields $\sigma$, $\alpha$, $\zeta$, and $\bar{\zeta}$ in
equation (\ref{SQ017}) we end up with
\begin{eqnarray}
Z\!\!&=&\!\!\!\int (\mathcal{D}S)(\mathcal{D}\lambda)(\mathcal{D}\psi)(\mathcal{D}\bar{\psi})
\delta\!\left(\!\sum_{{\bf r}}S_{{\bf r}}^2-N\!\right)
\delta\!\left(\!\sum_{{\bf r}}\bar{\psi}_{{\bf r}}S_{{\bf r}}\!\right)
\delta\!\left(\!\sum_{{\bf r}}\psi_{{\bf r}}S_{{\bf r}}\!\right)
\delta\!\left(\sum_{{\bf r}}
\lambda_{{\bf r}}S_{{\bf r}}-
\sum_{{\bf r}}\bar{\psi}_{{\bf r}}\psi_{{\bf r}}\!\!\right)
\nonumber\\
&\times&\exp\left[-\int dt\left(-\frac12\sum_{{\bf r}}\lambda_{{\bf r}}^2+
\sum_{{\bf r}}\lambda_{{\bf r}}\dot{S}_{{\bf r}}-
\sum_{{\bf r},{\bf r}'}U_{{\bf r},{\bf r}'}\lambda_{\bf r}S_{{\bf r}'}-\sum_{{\bf r}}
\bar{\psi}_{{\bf r}}\dot{\psi}_{{\bf r}}
+\sum_{{\bf r},{\bf r}'}U_{{\bf r},{\bf r}'}\bar{\psi}_{{\bf r}}
\psi_{{\bf r}'}
\right)\right].
\end{eqnarray}
In order to cast this expression in the desired final form, we need to perform a shift in the auxiliary field
in the functional integration $\lambda_{\bf r}\rightarrow\lambda_{\bf r}+\dot{S}_{\bf r}$
(that does not affect the constraint because
$\sum_{\bf r}\dot{S}_{\bf r}S_{\bf r}=\frac12\frac{d}{dt}(\sum_{\bf r}S_{\bf r}^2)=0$) and also it is
convenient to change $t\rightarrow-t$ in the action. After this, we get the partition function
\begin{eqnarray}
Z\!\!&=&\!\!\!\int (\mathcal{D}S)(\mathcal{D}\lambda)(\mathcal{D}\psi)(\mathcal{D}\bar{\psi})
\delta\!\left(\sum_{{\bf r}}S_{{\bf r}}^2-N\right)
\delta\!\left(\sum_{{\bf r}}\bar{\psi}_{{\bf r}}S_{{\bf r}}\right)
\delta\!\left(\sum_{{\bf r}}\psi_{{\bf r}}S_{{\bf r}}\right)
\delta\!\left(\sum_{{\bf r}}
\lambda_{{\bf r}}S_{{\bf r}}-
\sum_{{\bf r}}\bar{\psi}_{{\bf r}}\psi_{{\bf r}}\!\!\right)
\nonumber\\
&\times&\exp\left[-\int dt\left(-\frac12\sum_{{\bf r}}\lambda_{{\bf r}}^2+
\frac{1}{2}\sum_{{\bf r}}\dot{S}_{\bf r}^2
-\sum_{{\bf r},{\bf r}'}U_{{\bf r},{\bf r}'}\lambda_{\bf r}S_{{\bf r}'}+\sum_{{\bf r}}
\bar{\psi}_{{\bf r}}\dot{\psi}_{{\bf r}}
+\sum_{{\bf r},{\bf r}'}U_{{\bf r},{\bf r}'}\bar{\psi}_{{\bf r}}
\psi_{{\bf r}'}\right)\right].
\label{pf}
\end{eqnarray}
This expression is interesting due to the following reasons. First,
we may read out the Euclidean Lagrangian
\begin{eqnarray}
\mathcal{L}_E=-\frac12\sum_{{\bf r}}\lambda_{{\bf r}}^2+
\frac{1}{2}\sum_{{\bf r}}\dot{S}_{\bf r}^2
-\sum_{{\bf r},{\bf r}'}U_{{\bf r},{\bf r}'}\lambda_{\bf r}S_{{\bf r}'}+\sum_{{\bf r}}
\bar{\psi}_{{\bf r}}\dot{\psi}_{{\bf r}}
+\sum_{{\bf r},{\bf r}'}U_{{\bf r},{\bf r}'}\bar{\psi}_{{\bf r}}\psi_{{\bf r}'},
\label{lag}
\end{eqnarray}
that is invariant under the set of supersymmetry transformations
\begin{eqnarray}
\delta_{\epsilon}S_{{\bf r}}=\bar{\psi}_{{\bf r}}\epsilon,
~~~\delta_{\epsilon}\psi_{{\bf r}}=
\dot{S}_{{\bf r}}\epsilon+\lambda_{\bf r}\epsilon,~~~\delta_{\epsilon}\bar{\psi}_{{\bf r}}=0 ~~~\text{and}~~~
\delta_{\epsilon}\lambda_{\bf r}=-\dot{\bar\psi}_{\bf r}\epsilon
\label{r03}
\end{eqnarray}
and
\begin{eqnarray}
\delta_{\bar{\epsilon}}S_{{\bf r}}=
\bar{\epsilon}\psi_{{\bf r}}, ~~~
\delta_{\bar{\epsilon}}\psi_{{\bf r}}=0,~~~
\delta_{\bar{\epsilon}}\bar{\psi}_{{\bf r}}=
-\dot{S}_{{\bf r}}\bar{\epsilon}+\lambda_{\bf r}\bar\epsilon~~~\text{and}~~~
\delta_{\bar\epsilon}\lambda_{\bf r}=\bar\epsilon\dot{\psi}_{\bf r},
\label{r04}
\end{eqnarray}
where $\epsilon$ and $\bar{\epsilon}$ are fermionic infinitesimal parameters.
Moreover, the delta functions in the integrand impose the constraints
\begin{eqnarray}
\sum_{{\bf r}}S_{{\bf r}}^2=N,~~~\sum_{{\bf r}}\psi_{{\bf r}}S_{{\bf r}}=0,
~~~\sum_{{\bf r}}\bar{\psi}_{{\bf r}}S_{{\bf r}}=0,~~\text{and}~~
\sum_{\bf r}\lambda_{{\bf r}}S_{{\bf r}}
=\sum_{{\bf r}}\bar{\psi}_{{\bf r}}\psi_{{\bf r}},
\label{scc}
\end{eqnarray}
that turn out to be a supersymmetric constraints structure which are closed under (\ref{r03}) and (\ref{r04}).
From these observations, we recognize the model above obtained as an Euclidean off-shell
supersymmetric extension of the quantum spherical model. In this formulation, $\lambda_{\bf r}$ is the auxiliary field required by the
off-shell supersymmetry to guarantee the matching of the degrees of freedom.
We interpret the model defined from (\ref{pf}) as a quantum system constituted of bosonic ($S_{\bf r}$, $\lambda_{\bf r}$) and fermionic
($\psi_{\bf r}$, $\bar\psi_{\bf r}$) degrees of freedom at each lattice site,
behaving effectively in a supersymmetric way and satisfying the constraints (\ref{scc}).

To summarize, this supersymmetric version of the quantum spherical model arises from the
classical model by means of the stochastic quantization prescription. This comes from the construction
of a $D+1$-dimensional theory from a $D$-dimensional one (being careful to handle with the constraints),
with the properly identification of the fictitious time as the ordinary time and
exploring the supersymmetry in this time direction.

To show how the quantum spherical model (not supersymmetric) may be obtained from stochastic quantization scheme,
which is based in a procedure different from the above,
we study in the Appendix A the quantization of the mean spherical model along the lines of section \ref{section2a}.
There, the stochastic method is viewed strictly as a quantization method like the canonical or path integral quantization and, after we calculate the correlation function
and eliminate the fictitious time, we obtain the usual condition of the quantum spherical model \cite{Vojta1}.

The next step is to investigate both finite and zero temperature critical behaviors of the supersymmetric model.
For this, we need to calculate the partition function.


\section{Partition Function}

We will consider the large-$N$ limit (thermodynamic limit) evaluation of the partition function (\ref{pf}). We wish to take
into account the temperature, which means that the interval of integration of the Euclidean time runs from 0 to $\beta$ and furthermore the
fields satisfy specific periodic/anti-periodic conditions according to their bosonic/fermionic character: $S_{\bf r}(0)=S_{\bf r}(\beta)$,
$\lambda_{\bf r}(0)=\lambda_{\bf r}(\beta)$, $\psi_{\bf r}(0)=-\psi_{\bf r}(\beta)$, and $\bar\psi_{\bf r}(0)=-\bar\psi_{\bf r}(\beta)$.
In order to perform the Gaussian integrals we redefine the auxiliary field  $\lambda_{\bf r}\rightarrow i\lambda_{\bf r}$,
such that the path integral is given by
\begin{eqnarray}
&&Z\!\!=\!\!\!\int (\mathcal{D}S)(\mathcal{D}\lambda)(\mathcal{D}\psi)(\mathcal{D}\bar{\psi})
\delta\!\left(\sum_{{\bf r}}S_{{\bf r}}^2-N\right)
\delta\!\left(\sum_{{\bf r}}\bar{\psi}_{{\bf r}}S_{{\bf r}}\right)
\delta\!\left(\sum_{{\bf r}}\psi_{{\bf r}}S_{{\bf r}}\right)
\delta\!\left(\sum_{{\bf r}}
i\lambda_{{\bf r}}S_{{\bf r}}-
\sum_{{\bf r}}\bar{\psi}_{{\bf r}}\psi_{{\bf r}}\!\!\right)
\nonumber\\
&&\times\!\exp\left[-\!\int_{0}^{\beta}\! d\tau\left(\frac12\sum_{{\bf r}}\lambda_{{\bf r}}^2+
\frac{1}{2g}\sum_{{\bf r}}\dot{S}_{\bf r}^2
-i\sum_{{\bf r},{\bf r}'}U_{{\bf r},{\bf r}'}\lambda_{\bf r}S_{{\bf r}'}+\frac{1}{\sqrt{g}}\sum_{{\bf r}}
\bar{\psi}_{{\bf r}}\dot{\psi}_{{\bf r}}
+\!\sum_{{\bf r},{\bf r}'}U_{{\bf r},{\bf r}'}\bar{\psi}_{{\bf r}}
\psi_{{\bf r}'}\right)\right],
\label{5.1}
\end{eqnarray}
where we included a $g$-dependence in the kinetic terms that plays the role of a quantum coupling. Indeed,
 in the Hamiltonian formulation it will appear multiplying the momenta term, such that the limit $g\rightarrow 0$ will correspond formally to
the classical limit. 
The next step is to employ the functional integral representation  for the delta functions
\begin{equation}
\delta\big{(}\sum_{\bf r}S_{\bf r}^2-N\big{)}=\int\mathcal{D} \lambda\, e^{{-\int_{0}^{\beta}}d\tau\lambda\big{(}\sum_{\bf r}S_{\bf r}^2-N\big{)}},
\label{5.2}
\end{equation}
\begin{equation}
\delta\big{(}\sum_{\bf r}\bar{\psi}_{\bf r}S_{\bf r}\big{)}=\int \mathcal{D}\zeta \,e^{-\int_{0}^{\beta}d\tau\sum_{\bf r}\bar\psi_{\bf r}S_{\bf r}\zeta},
\label{5.3}
\end{equation}
\begin{equation}
\delta\big{(}\sum_{\bf r}\psi_{\bf r}S_{\bf r}\big{)}=\int \mathcal{D}\bar\zeta\, e^{-\int_{0}^{\beta}d\tau\sum_{\bf r}\bar\zeta\psi_{\bf r}S_{\bf r}},
\label{5.4}
\end{equation}
and
\begin{equation}
\delta\big{(}\sum_{\bf r}\bar{\psi}_{\bf r}{\psi}_{\bf r}-i\sum_{{\bf r}}S_{\bf r}\lambda_{{\bf r}}\big{)}=\int\mathcal{D} \gamma\, e^{{-\int_{0}^{\beta}}d\tau\gamma
\big{(} \sum_{\bf r}\bar{\psi}_{\bf r}{\psi}_{\bf r}-i\sum_{{\bf r}}S_{\bf r}\lambda_{{\bf r}} \big{)}}.
\label{5.5}
\end{equation}
We introduce the Fourier transformation of the fields depending on ${\bf r}$,
\begin{eqnarray}
(\cdots)_{\bf r}=\frac{1}{\sqrt{N}}\sum_{\bf q}
\text{e}^{i {\bf q}\cdot{\bf r}}(\cdots)_{\bf q},
\label{5.6}
\end{eqnarray}
with $(\cdots)_{\bf r}$ generically designating $S_{\bf r}$, $\lambda_{\bf r}$, $\psi_{\bf r}$, and $\bar{\psi}_{\bf r}$,
and $(\cdots)_{\bf q}$ the corresponding Fourier transformations.
The partition function becomes
\begin{eqnarray}
Z&=&\int (\mathcal{D}S)(\mathcal{D}\lambda)
(\mathcal{D}\psi)(\mathcal{D}\bar{\psi})
(\mathcal{D}\lambda)
(\mathcal{D}\gamma)(\mathcal{D}\xi)(\mathcal{D}\bar{\xi})\nonumber\\
&\times&\exp\left\{-\int_0^{\beta}d\tau\left[\frac{1}{2g}
\sum_{\bf q}\dot{S}_{\bf q}\dot{S}_{-\bf q}
+\frac{1}{2}\sum_{\bf q}\lambda_{\bf q}\lambda_{-\bf q}
-i\sum_{\bf q}{U}({\bf q})S_{\bf q}\lambda_{-\bf q}
+\lambda\left(\sum_{\bf q}{S}_{\bf q}{S}_{-\bf q}-N\right)
\right.\right.
\nonumber\\
&-&\left.\left.
i \gamma\sum_{\bf q}S_{\bf q}\lambda_{-\bf q}
+\sum_{\bf q}\bar{\psi}_{\bf q}\xi S_{-{\bf q}}
+\sum_{\bf q}\bar{\xi}\psi_{\bf q} S_{-{\bf q}}
+\sum_{\bf q}\bar{\psi}_{\bf q}\left(\frac{1}{\sqrt{g}}\frac{\partial}{\partial \tau}+
{U}({\bf q})+\gamma\right)\psi_{\bf q}\right]
\right\},
\label{5.7}
\end{eqnarray}
where we identified ${U}({\bf q})$ as the Fourier transformation of the interaction energy
$U_{{\bf r},{\bf r}'}$,
\begin{eqnarray}
{U}({\bf q})=\sum_{\bf h}U(|{\bf h}|)
\text{e}^{i{\bf q}\cdot{\bf h}},
~~~~\text{with}~~~~
{\bf h}={\bf r}-{\bf r}'.
\label{5.8}
\end{eqnarray}

Now we will perform the integrations over the fields $S_{\bf q}$, $\lambda_{\bf q}$, $\psi_{\bf q}$,
and $\bar{\psi}_{\bf q}$ to obtain the effective action. First, we consider the integration over the field $\lambda_{\bf q}$,
\begin{eqnarray}
\int\left(\mathcal{D}\lambda\right)
\exp\left\{-\int_0^{\beta}d\tau\left[\frac{1}{2}\sum_{\bf q}
\lambda_{\bf q}\lambda_{-{\bf q}}
-\frac{i}{2}\sum_{\bf q}
\left(\varphi_{\bf q}\lambda_{-{\bf q}}+
\varphi_{-{\bf q}}\lambda_{{\bf q}}\right)
\right]\right\},
\label{5.9}
\end{eqnarray}
with $\varphi_{\bf q}\equiv {U}({\bf q})S_{\bf q}+\gamma S_{\bf q}$. By using the
decompositions
\begin{eqnarray}
\lambda_{\bf q}=\text{Re}\lambda_{\bf q}
+i \text{Im}\lambda_{\bf q}
~~~~\text{and}~~~~
\varphi_{\bf q}=\text{Re}\varphi_{\bf q}
+i \text{Im}\varphi_{\bf q}
\label{5.10}
\end{eqnarray}
and the fact that $\varphi_{-{\bf q}}=\varphi^{\dag}_{\bf q}$ and
$\lambda_{-{\bf q}}=\lambda^{\dag}_{\bf q}$, the result of integral (\ref{5.9}) is
\begin{eqnarray}
\exp\left[
-\frac{1}{2}\int_0^{\beta}d\tau
\sum_{\bf q}\varphi_{\bf q}\varphi_{-{\bf q}}\right],
\label{5.11}
\end{eqnarray}
up to irrelevant constants.
Next, we consider the integral over $S_{\bf q}$. By
defining $\phi_{\bf q}\equiv(\bar{\psi}_{\bf q}\xi+\bar{\xi}\psi_{\bf q})$, and adopting a procedure similar to the above one, we have,
\begin{eqnarray}
&&\int(\mathcal{D}S)\exp\left\{
-\int_0^{\beta}d\tau\left[\sum_{\bf q}
S_{\bf q}\left(-\frac{1}{2g}
\frac{\partial^2}{\partial \tau^2}+
\lambda +\frac{1}{2}({U}({\bf q})
+\gamma)^2\right)S_{\bf q}
+\frac{1}{2}\sum_{\bf q}\left(\phi_{q}S_{-{\bf q}}
+\phi_{-{\bf q}}S_{\bf q}\right)\right]
\right\}\nonumber\\
&&=\exp\left(-\frac{1}{2}\text{Tr}\sum_{\bf q}\ln \mathcal{O}\right)
\exp\left(-\frac{1}{2}\int_0^{\beta}d\tau
\sum_{\bf q}\bar{\psi}_{\bf q}\xi\mathcal{O}^{-1}\bar{\xi}
\psi_{\bf q}\right),
\label{5.12}
\end{eqnarray}
where we defined the operator
\begin{equation}
\mathcal{O}\equiv-\frac{1}{2g}
\frac{\partial^2}{\partial \tau^2}+
\lambda +\frac{1}{2}({U}({\bf q})
+\gamma)^2,
\label{5.12a}
\end{equation}
and have used the identity $\det M=\exp(\text{Tr}\ln M)$.
Finally, we integrate over the $\psi_{\bf q}$ and $\bar{\psi}_{\bf q}$,
\begin{eqnarray}
&&\int(\mathcal{D}\bar{\psi})
(\mathcal{D}{\psi})\left[-\int_0^{\beta}
d\tau\sum_{\bf q}\bar{\psi}_{\bf q}\left(
\frac{1}{\sqrt{g}}\frac{\partial}{\partial \tau}
+{U}({\bf q})+\gamma+\frac{1}{2}\xi\mathcal{O}^{-1}
\bar{\xi}
\right)\psi_{\bf q}\right]\nonumber\\
&&=\det\left[\frac{1}{\sqrt{g}}\frac{\partial}
{\partial\tau}+{U}({\bf q})+\gamma
+\frac{1}{2}\xi\mathcal{O}^{-1}\bar{\xi}\right].
\label{5.13}
\end{eqnarray}
By collecting the above results, the partition function (\ref{5.7}) is given by
\begin{eqnarray}
Z=\int(\mathcal{D}\lambda)(\mathcal{D}\gamma)(\mathcal{D}\xi)(\mathcal{D}\bar{\xi})\exp\left(-NS_{eff}\right),
\label{5.14}
\end{eqnarray}
with the effective action
\begin{eqnarray}
S_{eff}&=&-\int_0^{\beta}
d\tau\lambda+\frac{1}{2N}\text{Tr}\sum_{\bf q}
\ln\left[-\frac{1}{2g}\frac{\partial^2}{\partial\tau^2}
+\lambda+\frac{1}{2}\left({U}({\bf q})+\gamma\right)^2
\right]\nonumber\\
&-&\frac{1}{N}\text{Tr}\sum_{\bf q}
\left[\frac{1}{\sqrt{g}}\frac{\partial}{\partial \tau}
+{U}({\bf q})+\gamma+\frac{1}{2}\bar{\xi}
\mathcal{O}^{-1}\xi\right].
\label{5.15}
\end{eqnarray}
In the calculation of the traces we need to take into account the different boundary conditions
for bosons and fermions in the Euclidean time.
The reflex of these conditions is the arising of the discrete spectrum for frequencies,
$\omega_n^B=2n\pi/\beta$ for the bosonic case and $\omega_n^F=(2n+1)\pi/\beta$ for the fermionic case, with $n\in \text{Z}$.

The saddle point equations are determined by means of the conditions
\begin{eqnarray}
\frac{\delta S_{eff}}{\delta \lambda}
=\frac{\delta S_{eff}}{\delta \gamma}
=\frac{\delta S_{eff}}{\delta \xi}
=\frac{\delta S_{eff}}{\delta \bar{\xi}}=0.
\label{5.16}
\end{eqnarray}
We shall look for solutions $(\lambda^{\ast},\gamma^{\ast},\xi^{\ast},\bar\xi^{\ast})$ with all these parameters independent of the time.
For simplicity of notation we write them simply as $(\lambda,\gamma,\xi,\bar\xi)$.
By using the identity $\delta\text{Tr}\ln A=\text{Tr}A^{-1}\delta A$ we find
that the conditions $\delta S_{eff}/\delta \xi=\delta S_{eff}/\delta \bar{\xi}=0$ are trivially satisfied with the choice
$\xi=\bar{\xi}=0$.
From the condition $\delta S_{eff}/\delta \lambda=0$, we obtain
\begin{eqnarray}
1=\frac{1}{N}\sum_{\bf q}
\frac{g}{2{w}_{\bf q}^B}\coth\left(
\frac{\beta {w_{\bf q}^B}}{2}\right),
\label{5.18}
\end{eqnarray}
with the bosonic frequency
\begin{eqnarray}
(w^B_{\bf q})^2\equiv 2g\left[\lambda
+\frac{1}{2}\left({U}({\bf q})
+\gamma\right)^2\right].
\label{5.19}
\end{eqnarray}
The last saddle point condition, $\delta S_{eff}/\delta \gamma=0$, leads to
\begin{eqnarray}
\frac{1}{N}\sum_{\bf q}\frac{g}{2w_{\bf q}^B}
\left({U}({\bf q})+\gamma\right)
\coth\left(\frac{\beta w_{\bf q}^B}{2}\right)
-\frac{1}{N}\sum_{\bf q}\frac{g}{2w_{\bf q}^F}
\left({U}({\bf q})+\gamma\right)
\tanh\left(\frac{\beta w_{\bf q}^B}{2}\right)=0,
\label{5.20}
\end{eqnarray}
where the fermionic frequency is defined as
\begin{eqnarray}
\left(w_{\bf q}^F\right)^2\equiv 2g\left[\frac{1}{2}
\left({U}({\bf q})+\gamma\right)^2\right].
\label{5.21}
\end{eqnarray}
The sum in ${\bf q}$ appearing in equations (\ref{5.18}) and (\ref{5.20}) must be understood as integrals
in the thermodynamic limit, $\frac{1}{N}\sum_{\bf q}\rightarrow \int d^Dq$. The critical behavior of the
system is determined by analyzing the saddle point equations near the critical point, that will be studied in Section \ref{section5}.

\subsection{Ground State Energy}

It is instructive to look at the ground state energy in order to make clear the situation
where the supersymmetry is not broken. This can be obtained from the free energy according to the standard relation
$f=-\frac{1}{\beta N}\ln Z$. By employing similar methods to the used above, we find the following result for the free energy
\begin{equation}
f=-\lambda+\frac{1}{\beta N}\sum_{\bf q}  \ln\left[2\sinh\left(\frac{\beta\omega_{\bf q}^B}{2}\right)\right]
-\frac{1}{\beta N}\sum_{\bf q}  \ln\left[2\cosh\left(\frac{\beta\omega_{\bf q}^F}{2}\right)\right],
\label{8.1}
\end{equation}
where the parameters $\lambda$ and $\gamma$ must satisfy (\ref{5.18}) and (\ref{5.20}).
Then by taking the zero temperature limit $T\rightarrow 0$ ($\beta\rightarrow\infty$) in this expression we obtain the
ground state energy $E_0$,
\begin{equation}
\frac{E_0}{N}=-\lambda+\frac{1}{N}\sum_{\bf q} (\omega_{\bf q}^B-\omega_{\bf q}^F).
\label{8.2}
\end{equation}
The supersymmetry requires the vanishing of the ground state energy. This is the case when
$\lambda=0$, where the bosonic and fermionic frequencies become equal $\omega_{\bf q}^B=\omega_{\bf q}^F$,
independently of the value of $\gamma$.  This show us that the requirement of supersymmetry, i.e., the
vanishing of the ground state energy, does not fix the parameter $\gamma$.

\section{Critical Behavior}\label{section5}

Let us investigate the critical behavior of the model by considering the saddle point conditions
(\ref{5.18}) and (\ref{5.20}) near the critical point. The critical point is obtained when $\lambda\rightarrow 0$ and $\gamma\rightarrow 0$,
and we parameterize the interaction $U({\bf q})$ in terms of a parameter $x$ as $U({\bf q})\sim q^{\frac{x}{2}}$ for small values of $q\equiv |{\bf q}|$.
We will also consider separately the situations of zero and nonzero temperatures.

We must restrict our analysis on two simple situations:

1. $\lambda=0$ and $\gamma\neq 0$,
where the bosonic and fermionic frequencies become equal, corresponding to the supersymmetric situation at zero temperature, as discussed before.
At the same time, as the parameter $\gamma$ can be interpreted as a mediator of the interaction between bosons and fermions,
a nonvanishing value of $\gamma$ implies a coupling between the bosonic and fermionic degrees of freedom;

2.  $\lambda\neq 0$ and $\gamma= 0$. This corresponds to the situation where the supersymmetry is explicitly broken.
Furthermore, in this situation there is a decoupling between the bosonic and fermionic degrees of freedom, because we
are \textquotedblleft turning off\textquotedblright the parameter responsible for implementation of the constraint that mix the corresponding degrees of freedom.

\subsection{Finite Temperature}

In this situation the thermal fluctuations are responsible to drive the phase transition and
the supersymmetry is broken by the temperature independently of the parameters $\lambda$ and $\gamma$, essentially
due to different distributions for bosons and fermions.
In order to study the critical behavior we expand the hyperbolic functions $\coth x=\frac{1}{x}+\frac{x}{3}+\cdots$
and $\tanh x=x+\cdots$, as the argument $x$ is small near the critical point.
The integrals in the equations (\ref{5.18}) and (\ref{5.20}) will converge if  $D>x$,
defining the lower critical dimension $D_l=x$.

\subsubsection{$\lambda=0$ and $\gamma\neq 0$}

By subtracting the equation (\ref{5.18}) near the critical point from itself at the critical point we get
\begin{equation}
t_g\,\,\,\sim\,\,\,\left\{
\begin{array}
[c]{cc}%
\gamma^{\frac{2(D-x)}{x}} & ~~ \left(  D<\frac{3x}{2}\right)  \\
\gamma\ln\gamma  & ~~\left( D=\frac{3x}{2}\right)\\
\gamma &~~\left( D>\frac{3x}{2}\right)
\end{array}
\right.,
\label{6.1}
\end{equation}
where $t_g\equiv (g-g_c)/g_c$, with $g_c$ being the value of $g$ at the critical point.
 We see that the upper critical dimension is $D_u=3x/2$, separating
nontrivial critical behavior ($D<D_u$) from mean-field behavior ($D>D_u$).
This result is different from that obtained for the quantum spherical model \cite{Vojta1} due to the
coupling between bosons and fermions, as argued before.


\subsubsection{$\lambda\neq 0$ and $\gamma= 0$}

In this situation, there is a decoupling between the bosonic and fermionic degrees of freedom and
we have effectively only the spherical constraint. Consequently, the critical
behavior reduces to that of the bosonic quantum spherical model \cite{Vojta1}
\begin{equation}
t_g\,\,\,\sim\,\,\,\left\{
\begin{array}
[c]{cc}%
\lambda^{\frac{D-x}{x}} & ~~ \left(  D<2x\right)  \\
\lambda\ln\lambda  & ~~\left( D=2x\right)\\
\lambda &~~\left( D>2x\right)
\end{array}
\right..
\label{6.2}
\end{equation}
The upper critical dimension is given by $D_u=2x$.

\subsection{Zero Temperature}

In this situation the phase transition is governed exclusively  by quantum fluctuations and it is really a quantum phase transition.
By taking the temperature equal to zero, the saddle point equations (\ref{5.18}) and (\ref{5.20}) reduce to
\begin{eqnarray}
1=\frac{1}{N}\sum_{\bf q}
\frac{g}{2{w}_{\bf q}^B}
\label{6.3}
\end{eqnarray}
and
\begin{eqnarray}
\frac{1}{N}\sum_{\bf q}\frac{g}{2w_{\bf q}^B}
\left({U}({\bf q})+\gamma\right)
-\frac{1}{N}\sum_{\bf q}\frac{g}{2w_{\bf q}^F}
\left({U}({\bf q})+\gamma\right)=0.
\label{6.4}
\end{eqnarray}
The sums above converge if $D>x/2$, defining the lower critical dimension $D_l=x/2$ in the quantum case.

\subsubsection{$\lambda=0$ and $\gamma\neq 0$}

This situation corresponds to the supersymmetric case, where the bosonic and fermionic frequencis are equal, $\omega_{\bf q}^B=\omega_{\bf q}^F$.
Note that the equation (\ref{6.4}) is identically satisfied.
The difference between the equation (\ref{6.3}) near the critical point and at the critical point yields to the following behavior
\begin{equation}
t_g\,\,\,\sim\,\,\,\left\{
\begin{array}
[c]{cc}%
\gamma^{\frac{2D-x}{x}} & ~~ \left(  D<x\right)  \\
\gamma\ln\gamma  & ~~\left( D=x\right)\\
\gamma &~~\left( D>x\right)
\end{array}
\right.,
\label{6.5}
\end{equation}
where $t_g\equiv (g^0-g_c^0)g_c^0$, with $g^0$ being the zero temperature value of $g$.
The upper critical dimension is $D_u=x$.
Thus we verify the existence of a quantum phase transitions even in the case where the supersymmetry is not broken, because
the saddle point conditions (\ref{6.3}) and (\ref{6.4}) as well as the vanishing of the ground state energy (\ref{8.2}) do not fix the value of $\gamma$.


\subsubsection{$\lambda\neq 0$ and $\gamma= 0$}

In the same way as in (\ref{6.2}) the critical behavior reduces to that of the bosonic quantum spherical model
\begin{equation}
t_g\,\,\,\sim\,\,\,\left\{
\begin{array}
[c]{cc}%
\lambda^{\frac{2D-x}{2x}} & ~~ \left(  D<\frac{3x}{2}\right)  \\
\lambda\ln\lambda  & ~~\left( D=\frac{3x}{2}\right)\\
\lambda &~~\left( D>\frac{3x}{2}\right)
\end{array}
\label{6.6}
\right..
\end{equation}
The quantum upper critical dimension is $D_u=3x/2$.


\section{Conclusions}

In this work, we reported some results about the stochastic quantization of the spherical model, bringing together methods from
statistical mechanics and field theory.
The two main results are the construction of the supersymmetric version of the quantum spherical model via stochastic quantization
and the analysis of its critical behavior, which shows that the model exhibits both classical and quantum phase transitions.

We started by reviewing some basic aspects of the method
with emphasis in the connection between the Langevin equation and the supersymmetric quantum mechanics,
aiming at the application of the corresponding  connection to the spherical model.
It is intuitive that the application to the spherical model leads to a systematic construction of a supersymmetric version.

On the basis of the Nicolai map, we  investigated the stochastic quantization of the strict spherical model. We showed
that the result of this process is an off-shell supersymmetric extension of the quantum spherical model
with the precise supersymmetric constraint structure.
This analysis establishes a connection between the classical model and its
supersymmetric quantum counterpart.
This supersymmetric construction is a more natural one and
gives further support and motivations to investigate  similar connections in other models of the literature.

We investigated the existence of classical and quantum phase transitions
in two particular situations according to different choices for the parameters $\lambda$ and $\gamma$, which
are the saddle point values of the fields that implement the constraints $\sum_{\bf r}S_{\bf r}^2-N=0$ and
$\sum_{\bf r}\bar\psi_{\bf r}\psi_{\bf r}-i\sum_{\bf r}S_{\bf r}\lambda_{\bf r}=0$, respectively.
In both situations the model exhibits classical and quantum phase transitions.
For $\lambda\neq 0$ and $\gamma=0$, corresponding to \textquotedblleft turning off\textquotedblright the constraint that couples the bosonic and
fermionic degrees of freedom, the critical behavior reduces effectively to that of the
bosonic quantum spherical model \cite{Vojta1}.
On the other hand, for $\lambda= 0$ and $\gamma\neq 0$, there is a coupling between the
bosonic and fermionic degrees of freedom, leading to a distinct critical behavior. At zero temperature, this
corresponds to the supersymmetric situation. We found a quantum phase transition even in this supersymmetric situation.
This result is different from that obtained with a previously constructed version in \cite{Gomes}.
There, a supersymmetric version of the quantum spherical model was constructed in {\it ad hoc} way,
without the presence of the auxiliary field $\lambda_{\bf r}$.
In the present work, we have constructed a supersymmetric model following the steps of the stochastic method
and have automatically obtained an off-shell version, involving the auxiliary field $\lambda_{\bf r}$.
Naively, we could think that the two models are the same 
after the elimination of the auxiliary field $\lambda_{\bf r}$ by integrating it out.
But, due to the presence of the constraints, which involve the auxiliary field, this process is not innocuous,
and will generate additional terms that are not present in the model in \cite{Gomes}.
Furthermore, in the present case, the vanishing of the ground state energy, required by supersymmetry,
does not fix the parameter $\gamma$, as it can be seen from equation (\ref{8.2}).


\section{Acknowledgments}

The authors thank Prof. S. R. Salinas for reading the manuscript, very useful discussions and criticism.
This work was partially supported by  Conselho
Nacional de Desenvolvimento Cient\'{\i}fico e Tecnol\'ogico (CNPq) and Funda\c{c}\~ao de Amparo a Pesquisa do Estado de S\~ao Paulo (FAPESP).


\appendix

\section{Stochastic Quantization of the Mean Spherical Model}

In this appendix, we wish to discuss the stochastic quantization of the mean spherical model
whose constraint is imposed as a thermal average $\sum_{\bf r}\langle S_{\bf r}^2\rangle=N$ (not strictly as in (\ref{SQ02b})) in contrast
with other approaches as the canonical and the  path integral quantization   methods
applied to this specific model \cite{Obermair,Vojta1,Chamati,Coutinho,Bienzobaz}. Let us consider the classical Hamiltonian
\begin{equation}
\mathcal{H}=\frac12 \,g \sum_{\bf r} P_{\bf r}^2
+\frac12\sum_{{\bf r},{\bf r}^{\prime}}J_{{\bf r},{\bf r}^{\prime}}S_{{\bf r}}S_{{\bf r}^{\prime}}+
\mu\left(\sum_{\bf r}S_{\bf r}^2-N\right),
\label{h1}
\end{equation}
where $g$ is a positive parameter associated with the dynamics. In fact,
the presence of the kinetic term is only to provide a dynamic to the model,
but we will not impose commutation relations between $S_{\bf r}$ and $P_{\bf r}$.
The quantum character comes from the stochastic process.
The corresponding Lagrangian in the Euclidean time (which includes a global signal redefinition) is given by
\begin{equation}
\mathcal{L}=\frac{1}{2g} \sum_{\bf r} \dot{S}_{\bf r}^2(t)
+\frac12\sum_{{\bf r},{\bf r}^{\prime}}J_{{\bf r},{\bf r}^{\prime}}S_{{\bf r}}(t)S_{{\bf r}^{\prime}}(t)+
\mu\left(\sum_{\bf r}S_{\bf r}^2(t)-N\right).
\label{l1}
\end{equation}
We shall use the variable $t$ to represent the Euclidean time and the variable $\tau$ will be reserved to  the fictitious time in the stochastic quantization process. We start the stochastic quantization procedure by assuming that the spin variables
depend on the fictitious time $S_{\bf r}(t)\rightarrow S_{\bf r}(t,\tau)$, whose evolution is governed by the Langevin equation
(with $\kappa=2$),
\begin{equation}
\frac{\partial S_{{\bf r}}(t,\tau)}{\partial \tau}
= -\frac{\delta S}{\delta S_{{\bf r}}(t,\tau)}
+\eta_{{\bf r}}(t,\tau),
\end{equation}
where $S$ is the Euclidean action,
\begin{equation}
S=\int dt d\tau \left[\frac{1}{2g} \sum_{\bf r} \dot{S}_{\bf r}^2(t,\tau)
+\frac12\sum_{{\bf r},{\bf r}^{\prime}}J_{{\bf r},{\bf r}^{\prime}}S_{{\bf r}}(t,\tau)S_{{\bf r}^{\prime}}(t,\tau)+
\mu\left(\sum_{\bf r}S_{\bf r}^2(t,\tau)-N\right)\right],
\end{equation}
and $\eta_{{\bf r}}(t,\tau)$ is the Gaussian noise. It is worth to stress that the Langevin dynamics considered here
constitutes just an intermediate step towards the quantization of the model.
Notice that the Langevin dynamics itself for the classical spherical model was studied by several authors,
for example \cite{Luck,Stariolo,Henkel,Hase}.

We will consider the problem of stochastic quantization at finite temperature,
that can be put in consonance with the imaginary time formalism of field theory. In this situation,
the Euclidean or imaginary time is restricted to the interval $[0,\beta]$ and the fields must satisfy
periodic or anti-periodic conditions according to their bosonic or fermionic character, respectively, i.e.,
$\varphi(0)=\pm\varphi(\beta)$. Besides the fields, the noise
must also satisfy periodic or anti-periodic conditions. Namely, for a bosonic or fermionic noise we have
$\eta(0)=\pm\eta(\beta)$, assigning to it the same character as the corresponding field.
Then, the Gaussian noise average must be
adequate to reflect the periodic conditions of the bosonic fields in the imaginary time \cite{Namiki1}, i.e.,
\begin{equation}
\left<\eta_{{\bf r}}(t,\tau)\right>=0~~~\text{and}~~~
\left<\eta_{{\bf r}}(t,\tau)\eta_{{\bf r}'}(t',\tau')\right>=\frac{2}{\beta}\,
\sum_{n=-\infty}^{\infty}\text{e}^{i\omega_n(t-t')}\delta_{{\bf r},{\bf r}'}\delta(\tau-\tau'),
\label{momentum}
\end{equation}
where $\omega_n$ is the Matsubara frequency $\omega_n\equiv2\pi n/\beta$, with $n\in Z$.
It is convenient to introduce the Fourier decomposition of the variables $S_{{\bf r}}(t,\tau)$
and $\eta_{{\bf r}}(t,\tau)$ taking into account the periodic conditions,
\begin{equation}
S_{{\bf r}}(t,\tau)=\left(\frac{\beta}{N}\right)^{\frac{1}{2}}
\sum_{{\bf q},n} \text{e}^{i\left({\bf r}\cdot{\bf q}+\omega_nt\right)}
{S}_{{\bf q}}(\omega_n,\tau)~~~~\text{and}
~~~~
\eta_{{\bf r}}(t,\tau)=\left(\frac{\beta}{N}\right)^{\frac{1}{2}}
\sum_{{\bf q},n} \text{e}^{i\left({\bf r}\cdot{\bf q}+\omega_nt\right)}\,{\eta}_{{\bf q}}(\omega_n,\tau).
\label{fe}
\end{equation}
By identifying ${J}({\bf q})$ as the Fourier transformation of the interaction energy $J_{{\bf r},{\bf r}'}$,
\begin{equation}
{J}({\bf q})=\sum_{{\bf h}}J(|{\bf h}|)\,\text{e}^{-i{\bf q}\cdot{\bf h}},
\end{equation}
with ${\bf h}={\bf r}-{\bf r}'$, the Langevin equation in the Fourier space takes the form
\begin{eqnarray}
\frac{\partial {S}_{\bf q}(\omega_n,\tau)}{\partial \tau}
+\left[\frac{\omega_n^2}{g}+2\mu+{J}({\bf q})\right]{S}_{\bf q}(\omega_n,\tau)
-{\eta}_{{\bf q}}(\omega_n,\tau)=0,
\end{eqnarray}
whose general solution is given by
\begin{equation}
{S}_{\bf q}(\omega_n,\tau)={S}_{\bf q}(\omega_n,0)\,\text{e}^{-\Omega^2({\bf q},n)\tau}
+\int_0^{\tau}d\tilde{\tau}\,\text{e}^{-\Omega^2({\bf q},n)(\tau-\tilde{\tau})}
\,{\eta}_{{\bf q}}(\omega_n,\tilde{\tau}),
\label{Sq}
\end{equation}
where $\Omega({\bf q},n)$ is defined as
\begin{equation}
\Omega^2({\bf q},n)\equiv\frac{1}{g}
\left[\omega_n^2+2g\left(\mu+\frac{{J}({\bf q})}{2}\right)\right].
\end{equation}
From the solution (\ref{Sq}) we can determine the correlation functions of the mean spherical model according to (\ref{momentum}).
As stated early, the correlation functions so obtained must recover those of the quantized spherical model
whenever we take the appropriate limit of the $n$-point function $\tau_1=\tau_2=\cdots=\tau_n\equiv\tau\rightarrow \infty$.
In particular, for the 2-point function we get
\begin{eqnarray}
\left<S_{{\bf q}}(\omega_n,\tau)S_{{\bf q}'}(\omega_{n'},\tau')\right>
&=&\frac{1}{\beta^2\Omega^2({\bf q},n)}\,
\delta_{{\bf q},-{\bf q}'}\delta_{n,-n'}
\text{e}^{-\Omega^2({\bf q},n)|\tau-\tau'|}.
\label{momentum1}
\end{eqnarray}
In this expression we have discarded terms that do not contribute in the equilibrium limit.
Then, by taking $\tau=\tau'\rightarrow \infty$, we find the correlation function in the momentum  space for the
quantum spherical model
\begin{eqnarray}
\left<S_{{\bf q}}(\omega_n)S_{{\bf q}'}(\omega_n')\right>
= \delta_{{\bf q},-{\bf q}'}\delta_{n,-n'}\frac{1}{\beta^2}
\frac{g}{\left[\omega_n^2+w^2_{\bf q}\right]},
\end{eqnarray}
where
\begin{eqnarray}
w^2_{\bf q}\equiv2g\left(\mu
+\frac{J({\bf q})}{2}\right).
\end{eqnarray}
Consequently, the correlation function in the coordinate space is given by
\begin{eqnarray}
\left<S_{{\bf r}}(t)S_{{\bf r}'}(t')\right>
&=&\frac{1}{N}\sum_{{\bf q}}\frac{g}{2w_{\bf q}}
\frac{\text{e}^{- (t-t')w_{\bf q}}+\text{e}^{- [\beta-(t-t')]w_{\bf q}}}{1-\text{e}^{-\beta w_{\bf q}}}\,\text{e}^{i{\bf q}\cdot({\bf r}-{\bf r}')}.
\end{eqnarray}
The equal-time correlation function reduces to,
\begin{eqnarray}
\left<S_{{\bf r}}S_{{\bf r}'}\right>=
\frac{1}{N}\sum_{{\bf q}}\frac{g}{2w_{\bf q}}
\coth\left(\frac{\beta w_{\bf q}}{2}\right)\text{e}^{i{\bf q}\cdot({\bf r}-{\bf r}')},
\label{qcf}
\end{eqnarray}
that is the usual quantum pair correlation function of the spherical model (see, for example \cite{Coutinho}).
We can also obtain the classical correlation function by taking the limit $g\rightarrow 0$:
\begin{eqnarray}
\left<S_{\bf r}S_{{\bf r}'}\right>
=\frac{1}{N}\sum_{{\bf q}}\frac{1}{2\beta}\frac{1}{\left(\mu
+\frac{J({\bf q})}{2}\right)}~
\text{e}^{i{\bf q}\cdot({\bf r}-{\bf r}')}.
\end{eqnarray}
The quantum self-correlation furnishes the constraint condition corresponding
to the mean spherical model if we impose $\sum_{\bf r}\langle S_{\bf r}^2\rangle=N$,
\begin{eqnarray}
1-\frac{1}{N}\sum_{{\bf q}}\frac{g}{2w_{\bf q}}\coth\left(\frac{\beta w_{\bf q}}{2}\right)=0.
\label{qqq}
\end{eqnarray}
The  constraint equation (\ref{qqq}) may be used to determine the critical behavior of the model \cite{Vojta1,Gomes}.
To this purpose it is convenient to parametrize the interaction energy as $J({\bf q})\sim |{\bf q}|^x$ for
small values of $|{\bf q}|$. The parameter $x$ specifies the range of interactions, for example, for short-range interactions we have $x=2$.
The quantized spherical model is quite interesting in the sense that it exhibits a quantum phase transition, i.e., a phase
transition at zero temperature besides the usual finite temperature phase transition.


\end{document}